\documentclass[12pt, twocolumn]{openjournal}
\usepackage{graphicx,float}
\usepackage{xcolor}
\usepackage{hyperref}
\hypersetup{
    colorlinks=true,
    linkcolor=blue,
    filecolor=blue,      
    urlcolor=blue,
    citecolor=blue,
}
\usepackage{color,colortbl}
\DeclareGraphicsExtensions{.bmp,.png,.jpg,.pdf}
\usepackage{orcidlink}
\usepackage{amsmath}
\usepackage{framed}
\begin{document}
\title{Second Order Closures for the Radiative Transfer Equation: Some Are Unstable \vspace{-15mm}}
\author{Nickolay Y. Gnedin$^{1,2,3}$\orcidlink{0000-0001-5925-4580} and Harley Katz$^{2,3}$}
\affiliation{$^1$ Theory Division; 
Fermi National Accelerator Laboratory;
Batavia, IL 60510}
\affiliation{$^2$ Department of Astronomy \& Astrophysics; 
The University of Chicago; 
Chicago, IL 60637}
\affiliation{$^3$ Kavli Institute for Cosmological Physics;
The University of Chicago;
Chicago, IL 60637, USA}

\begin{abstract}
The largest existing simulations of cosmic reionization model radiative transfer with moment methods that require a closure relation. The two most commonly used closure relations are M1 and OTVET; both close the moment hierarchy at the first moment. We explore the properties of a higher, second-order closure. We show that direct generalizations of M1 and OTVET to one higher order are physically unstable - i.e., the closure equations themselves result in unstable solutions, not just their numerical implementation. In fact, a generalization of OTVET to any order higher than the first one is unstable. We are also able to show that any local (i.e., depending only on the local moments of the radiation field, like M1) second-order closure that depends only on the radiation intensity and radiation flux, but does not explicitly depend on the radiation pressure, is physically unstable. This result restricts the choice of possible second-order closure relations.
\end{abstract}


\maketitle

\section{Introduction}

Radiative transfer is a key ingredient in many astrophysical systems, from stellar atmospheres and radiation-mediated shocks to star formation, accretion flows, galaxy evolution, and cosmic reionization. In most multidimensional applications, directly solving the time-dependent transfer equation for the specific intensity $I_\nu(\vec{x},\vec{n},t)$ is prohibitively expensive because of its high dimensionality. A common alternative is to evolve angular moments of the radiation field and to truncate the resulting hierarchy by prescribing a closure relation for the highest retained moment.

Taking angular moments of the transfer equation yields evolution equations for the radiation energy density $E_\nu$, flux $F_\nu^i$, and, optionally, the radiation pressure $P_\nu^{ij}$ and higher-order tensors. Truncation at a finite order requires an algebraic or nonlocal relationship between successive moments. In the diffusion (Eddington) approximations, one effectively assumes near-isotropy and closes the system with the radiation pressure tensor $P_\nu^{ij}\approx (E_\nu/3)\delta^{ij}$, while flux-limited diffusion (FLD) modifies the diffusive flux to enforce causality in optically thin regions \citep{LevermorePomraning1981,Levermore1984}. Variable Eddington tensor (VET) approaches instead evolve a subset of the moment equations while determining an Eddington tensor from auxiliary considerations, ranging from approximate analytic forms to more sophisticated full radiative transfer solvers \citep{Finlator2009,Menon2022}.

A widely used hyperbolic two-moment formulation in astrophysics is the so-called ``M1'' model \citep{Levermore1984}, in which the moment hierarchy for the radiation energy density and flux is closed by approximating the radiation pressure tensor $P_\nu^{ij}$ as a local function of the radiation energy density and flux. The maximum-entropy closure similarly approximates the Eddington tensor as a local function of $E_\nu$ and $F_\nu^i$, with a slightly different functional form \citep{Minerbo1978,Levermore1984}. This approach has enabled robust and scalable radiation-(magneto)hydrodynamics implementations across a wide range of problems, including supernova and star-formation contexts (e.g.\ \citealt{GonzalezAuditHuynh2007}) and cosmological radiation hydrodynamics (e.g.\ \citealt{Rosdahl2013,RosdahlTeyssier2015,Kannan2019}). Related analytic closures and comparisons, including those motivated by neutrino transport, further highlight how sensitive multidimensional behavior can be to the adopted closure \citep{Murchikova2017}.

In cosmology, the ``Optically Thin Variable Eddington Tensor'' (OTVET) method computes the Eddington tensor in the optically thin limit, enabling a closure whose cost is essentially independent of the number of sources \citep{Gnedin2001}. OTVET and related moment-based VET approaches have been incorporated into cosmological radiative transfer and radiation-hydrodynamics frameworks and compared against other cosmological RT methods \citep{IlievMellemaPenEtAl2006,PetkovaSpringel2009,Gnedin2014}.

Recent comparisons between large cosmological simulations of reionization emphasized the limitations of the first moment closures. M1-based Thesan \citep{Kannan2022} and OTVET-based CROC \citep[``Cosmic Reionization on Computers''][]{Gnedin2014} show a surprising degree of difference in the evolution of the radiation field, while their radiative sources appear to evolve and cluster similarly \citep{Gnedin2025}. This difference serves as a motivation to increase the accuracy of the cosmological radiative transfer, and one pathway for doing so is to increase the order of the moment closure. 

The mathematical structure of truncated moment systems depends crucially on the closure. Questions of hyperbolicity, realizability, characteristic speeds, and stability are central both for continuum well-posedness and for the behavior of numerical schemes. Foundational work by Levermore and collaborators clarified the relationship between variable Eddington factors and flux limiters and established widely used FLD/VET constructions \citep{LevermorePomraning1981,Levermore1984}. For entropy-based $M_1$ models, rigorous analyses have addressed well-posedness and asymptotic limits \citep{GoudonLin2013}. In astrophysical settings, the characteristic structure of two-moment systems with nonlinear closures has also been examined, including the appearance of critical points in spherically symmetric configurations \citep{SmitCernohorskyDullemond1997}. 

This work builds on that tradition by focusing on the mathematical properties of second-order moment closures used in astrophysics, with emphasis on their stability and internal consistency (such as being consistent with the $1/r^2$ radiation field of a point source.

\section{Moment Hierarchy Closure Primer}

There appears to exist some confusion about the nomenclature of moment hierarchy closures, so we rehash them here for completeness. The monochromatic specific intensity $I_\nu(\vec{x},\vec{n},t)$ is a function of 6 phase-space variables plus time. Since solving a radiative transfer equation for $I_\nu$ in 6D phase space is a formidable numerical challenge, a single 6D+time equation is often replaced with a hierarchy of 3D+time equations for its moments. The first four moments of $I_\nu$ are
\begin{subequations}%
\label{eq:momdef}%
\begin{align}
    E_\nu(\vec{x},t) & = \frac{1}{c} \int d\Omega\, I_\nu(\vec{x},\vec{n},t), \\
    M_\nu^i(\vec{x},t) & = \frac{1}{c} \int d\Omega\, I_\nu(\vec{x},\vec{n},t) n^i = \langle n^i\rangle E_\nu, \\
    P_\nu^{ij}(\vec{x},t) & = \frac{1}{c} \int d\Omega\, I_\nu(\vec{x},\vec{n},t) n^i n^j = \langle n^i n^j\rangle E_\nu, \\
    Q_\nu^{ijk}(\vec{x},t) & = \frac{1}{c} \int d\Omega\, I_\nu(\vec{x},\vec{n},t) n^i n^j n^k = \langle n^i n^j n^k\rangle E_\nu, 
\end{align}
\end{subequations}
where $E_\nu$ is the photon energy density, $F_\nu^i \equiv c M_\nu^i$ is the photon flux, and $P_\nu^{ij}$ is the photon pressure tensor. In the absence of a better name, we call the third moment $Q_\nu^{ijk}$ the ``heat tensor''. 

In the following, we omit the frequency subscript for clearer notation - all radiation field quantities are henceforth implicitly frequency-dependent. With this notation, the equations for the first 3 moments are:
\begin{subequations}%
\label{eq:moms}%
\begin{align}
    \frac{\partial E}{c\partial t}+ \frac{\partial M^j}{\partial x^j} & = 
    -\kappa E + \frac{\dot{E}}{c}, \\
    \frac{\partial M^i}{c\partial t} + \frac{\partial P^{ij}}{\partial x^j} & = 
    -\kappa M^i + \frac{\dot{M}^i}{c}, \\
    \frac{\partial P^{ij}}{c\partial t} + \frac{\partial Q^{ijk}}{\partial x^k} & = 
    -\kappa P^{ij} + \frac{\dot{P}^{ij}}{c}, 
\end{align}
\end{subequations}
where $\dot{E}$ is the photon production rate density (i.e., luminosity density), $c\dot{M}^i$ is the flux source, and $\dot{P}^{ij}$ is the photon pressure source. For isotropic sources $\dot{M}^i = 0$ and $\dot{P}^{ij} = \dot{E}\delta^{ij}/3$.

The moment hierarchy is, generally, not closed (i.e., solutions for the lower moments require knowledge of a higher moment), and hence a closure relation is required. Any closure is an ansatz, and the resultant truncated hierarchy is (in a general sense) only an approximation. The simplest (and historically by far the most common) closure truncates the hierarchy at the first moment ($M^i$ - moments are counted starting at zero) equation. Such a closure is generally known as a ``Variable Eddington Tensor'' approximation. The radiation pressure tensor $P^{ij}= \langle n^i n^j\rangle E$ has a general property that its trace is equal to $E$, hence the Eddington tensor $h^{ij}$, defined as
\[
    P^{ij} = E h^{ij},
\]
is a symmetric, positive definite tensor with unit trace.

One common class of closures is \emph{local closures}, in which the highest-order moment (radiation pressure for the 1st-order closure, the heat tensor for the 2nd-order closures, etc) depends only on the local values of other radiation moments. For example, in the 1st-order closure, the radiation pressure tensor $P^{ij}$ and the Eddington tensor $h^{ij}$ depend on the values of $E$ and $M^i$ at the same spatial location only. The most general form of a symmetric, positive definite tensor with unit trace that depends only on the local properties of the radiation field is
\begin{equation}
    h^{ij} = A(f) \frac{M^i M^j}{M^2} + \frac{1-A(f)}{3}\delta^{ij},
    \label{eq:m1}
\end{equation}
where $f \equiv |\vec{M}|/E$.  The main feature of local closures is that $A(1)=1$ and $A(0)=0$, so the closure smoothly transitions from a ``streaming'' limit (radiation propagates along a single direction $\vec{u}$, $M^i=Eu^i$, $P^{ij}=E u^i u^j$) at $f=1$ to a diffusion limit ($P^{ij}=E \delta^{ij}/3$ at $f=0$).

A well-known example of a local 1st-order closure is ``M1'' \citep{Levermore1984}. In the M1 ansatz, the radiation field is assumed to be isotropic in some other reference frame (labeled by subscript 0, $M_0^i=0$, $P_0^{ij}=E_0\delta^{ij}/3$), while in the reference frame under consideration (the frame of Equation (\ref{eq:m1})), the radiation moments are Lorentz-boosted by some boost factor $\beta$:
\begin{align}
    E & = \gamma^2 E_0(1+\beta^2/3),\nonumber\\
    M^i & =\frac{4}{3}\gamma^2E_0\beta^i.\nonumber
\end{align}
Solving for $\beta^i$ from these equations and computing the Eddington tensor in the boosted reference frame results in Equation (\ref{eq:m1}) with 
\[
    A(f) = \frac{6f^2+2-\sqrt{4-3f^2}}{5+2\sqrt{4-3f^2}}.
\]

The M1 closure can be straightforwardly extended to the second order. In the Lorentz-boosted reference frame the heat tensor $Q^{ijk}$ depends only on $E$ and $M^i$, hence the most general form of such a tensor is
\begin{align}
    Q^{ijk} & = A_2\frac{M^iM^jM^k}{M^2} \nonumber\\
              & + \frac{1-A_2}{5}\left(M^i\delta^{jk}+M^j\delta^{ik}+M^k\delta^{ij}\right),
    \label{eq:m2}
\end{align}
where $A_2$ is a function of $f$. It can be computed in a closed form, but we do not write it out here explicitly because, as we show below, this closure is unstable. 

Another commonly used closure is called ``Maximum Entropy'' \citep{Minerbo1978,Levermore1984}. In the Maximum Entropy closure, the full radiation field is approximated as
\begin{equation}
    I(\vec{x},\vec{n},t) = I_0 \exp\left(a+b_in^i + c_{ij}n^i n^j + ...\right),
    \label{eq:me}
\end{equation}
where $I_0$ is a dimensional constant, and $a$, $b_i$, $c_{ij}$, etc are functions of $\vec{x}$ and $t$. For the first-order closure, $c_{ij}$ and other higher-order coefficients are set to zero, so that equations (\ref{eq:momdef}a,b) for the first two moments become equations for $a$ and $b_i$ as functions of $E$ and $M^i$,
\begin{align}
    E & = \frac{I_0}{c} \int e^{a+b_i n^i} d\Omega,\nonumber\\
    M^i & =\frac{I_0}{c} \int n^i e^{a+b_i n^i} d\Omega.\nonumber
\end{align}
The radiation pressure tensor is then given by
\[
  P^{ij} =\frac{I_0}{c} \int n^i n^j e^{a+b_i n^i} d\Omega.
\]

\begin{figure}[t]
\centering
\includegraphics[width=\columnwidth]{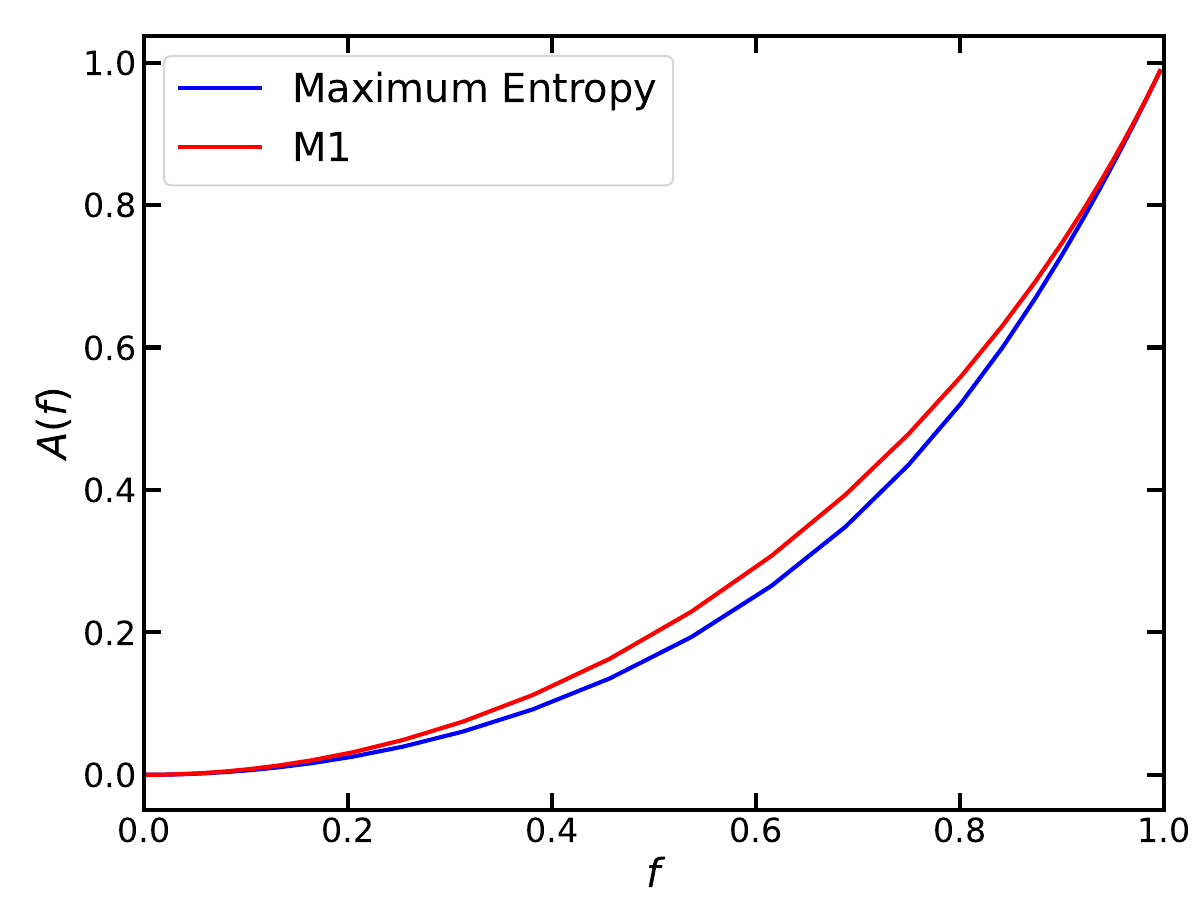}%
\caption{$A(f)$ versus $f$ for M1 and Maximum Entropy closures.}
\label{fig:am1me}
\end{figure}

These integrals can be taken analytically, resulting in vector $\vec{b}_i$ being aligned with vector $M^i$, and its amplitude $b$ being a solution of the transcendental algebraic equation
\[
    f = -\frac{1}{b} + \coth(b),
\]
where $f=M/E$ as before. The Eddington tensor for the first-order Maximum Entropy closure has the same form (\ref{eq:m1}) as the M1 closure, but the function $A$ is different, $A=1-f/b$. Figure \ref{fig:am1me} compares functions $A(f)$ for M1 and Maximum Entropy closures. They are similar, but not identical - Maximum Entropy results in a slightly more isotropic radiation field than M1.

The Maximum Entropy closure can analogously be extended to the second (or any other) order. In the second order, $c_{ij}$ is no longer assumed to be zero, with $a$, $b_i$, and $c_{ij}$ being solutions to the following 3 equations:
\begin{subequations}%
\label{eq:me2}%
\begin{align}
    E & = \frac{I_0}{c} \int e^{a+b_i n^i+c_{ij}n^in^j} d\Omega,\\
    M^i & =\frac{I_0}{c} \int n^i e^{a+b_i n^i+c_{ij}n^in^j} d\Omega,\\
    P^{ij} & =\frac{I_0}{c} \int n^i n^j e^{a+b_i n^i+c_{ij}n^in^j} d\Omega.
\end{align}
\end{subequations}
These integrals cannot be taken analytically, so the solution of the second-order Maximum Entropy closure cannot be presented in a closed form. Having solved for $a$, $b_i$, and $c_{ij}$ numerically, one can then close the moment hierarchy with the heat tensor
\begin{equation}
    Q^{ijk} = \frac{I_0}{c} \int n^i n^j n^k e^{a+b_i n^i+c_{ij}n^in^j} d\Omega.
\label{eq:me2clos}%
\end{equation}

Unfortunately, there is substantial confusion in the literature about the nomenclature. The label ``M$_N$'' is often used to refer to higher-order Maximum Entropy closures rather than higher-order analogs of M1 \citep[e.g.][]{DubrocaFeugeas1999,hauck2011,alldredge2013,guisset2015,pichard2016}, and even the label ``M1'' is sometimes used to refer to the first-order Maximum Entropy closure rather than the actual M1 closure \citep[e.g.][]{guisset2015}. Hence, to reduce confusion, we avoid using the label ``M2'' altogether and explicitly refer to the ``second-order M1-like scheme'' for the closure that computes the heat tensor in the Lorenz-boosted frame (Eq.\ \ref{eq:m2}) and the ``second-order Maximum Entropy scheme'' for the closure from Equation (\ref{eq:me2clos}).

The OTVET closure \citep{Gnedin2001} takes a different approach and evaluates the Eddington tensor in the optically thin limit,
\[
    h^{ij} = \frac{1}{4\pi c E_{\rm OT}} \int d^3 x^\prime \frac{\dot{E}(\vec{x}^\prime)}{(\vec{x}-\vec{x}^\prime)^2} u^i u^j,
\]
where
\[
    u^i \equiv \frac{x^i-x^{\prime i}}{|\vec{x}-\vec{x}^\prime|}
\]
and
\begin{equation}
    E_{\rm OT} = \frac{1}{4\pi c} \int d^3 x^\prime \frac{\dot{E}(\vec{x}^\prime)}{(\vec{x}-\vec{x}^\prime)^2}.
    \label{eq:eot}    
\end{equation}

An important requirement for any closure scheme is the ability to recover analytical solutions for relevant test cases. In particular, unless the numerical scheme is used exclusively in the optically thick regime, the closure should be able to support an optically thin single source solution,
\[
    E(\vec{x}) = \frac{L}{4\pi c r^2}.
\]
Both local closures and OTVET satisfy this requirement: local closures by virtue of the condition $A(1)=1$ and OTVET by construction, since it is exact in the optically thin limit (OTVET is also exact for a single source with an arbitrary opacity field $\kappa(\vec{x})$).

\begin{figure}[t]
\centering
\includegraphics[width=\columnwidth]{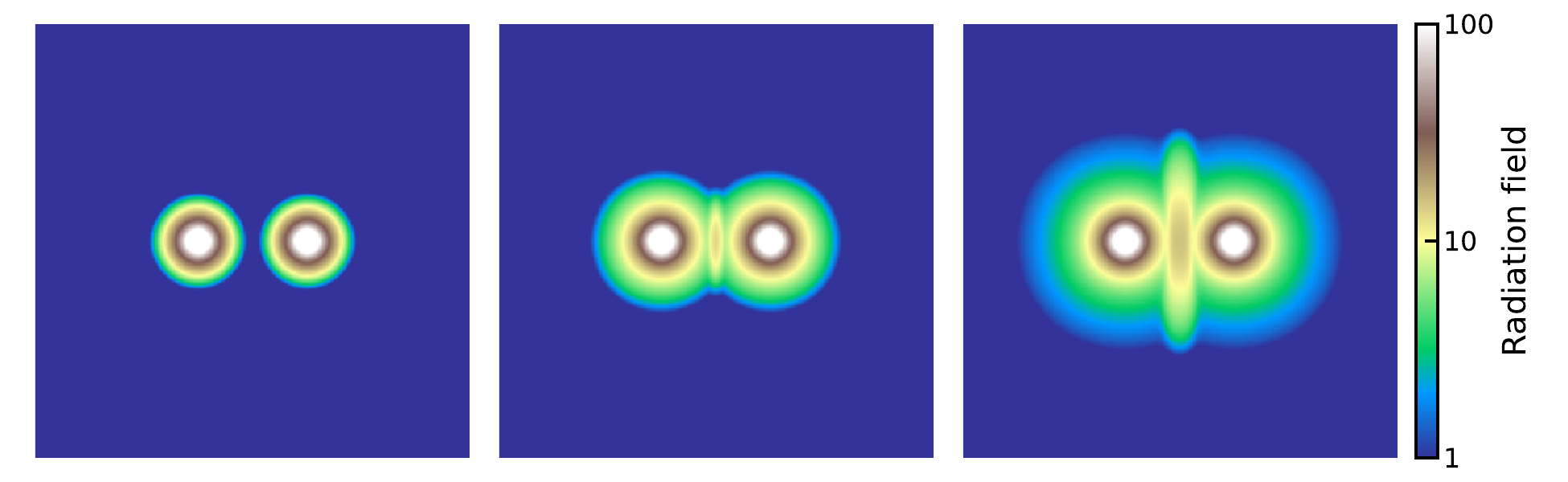}\newline%
\includegraphics[width=\columnwidth]{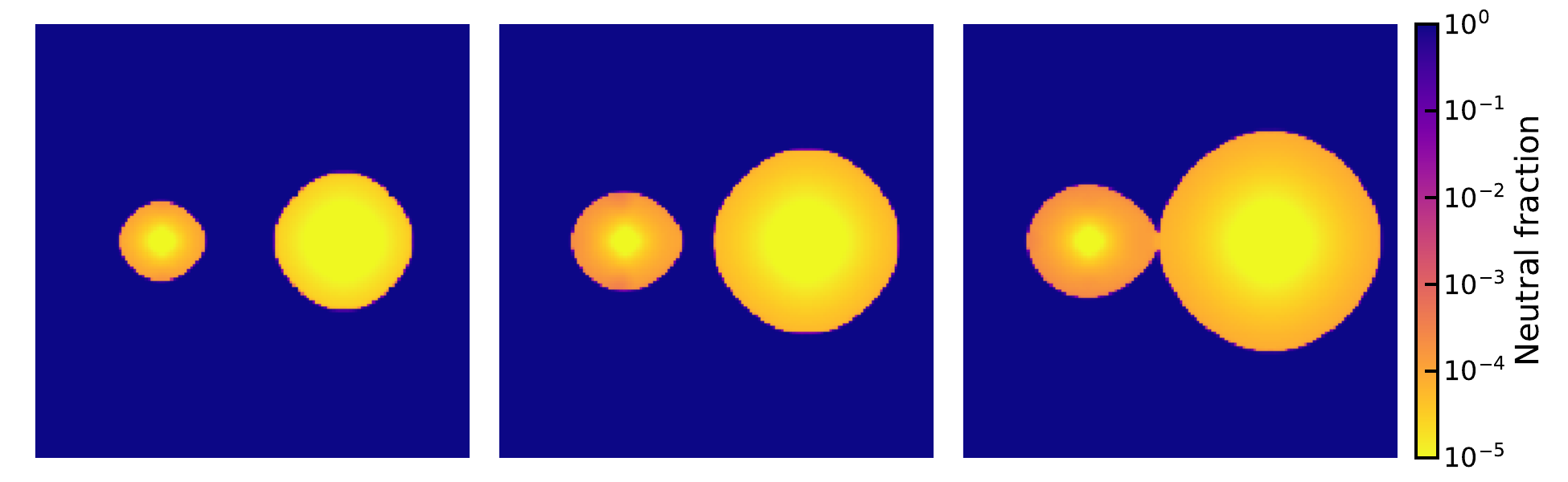}%
\caption{Top: radiation field in M1 closure (in arbitrary units) from two equal-luminosity sources at three different snapshots. Bottom: ionized bubbles in OTVET closure around two unequal sources; the right source is 10 times more luminous than the left one. In both rows, there are noticeable artifacts.}
\label{fig:arts}
\end{figure}

\section{Well-known Problems with First Order Closures}

However, no closure relation is exact in general. Thus, both M1 and OTVET exhibit well-known artifacts. These are already evident in a simple test of two radiation sources, as shown in Figure \ref{fig:arts}. These and all subsequent tests are performed with the simple global Lax-Friedrich scheme \citep{Aubert2008,Rosdahl2013} to calculate inter-cell fluxes, using the specific numerical discretization scheme explicitly described in \citet{Gnedin2026}\footnote{The results are insensitive to this choice.}. In this test, the computational volume is set to 2 Mpc, and the test is evolved for 20 Myr, ensuring that the light front originating at the box center does not reach the box edge by the end of the test.

In the M1 test, the space is assumed empty (and hence the frequency of the radiation is irrelevant). Since at the exact midpoint the total flux is formally zero (fluxes from left and right cancel exactly), the Eddington tensor becomes isotropic, and the radiation starts streaming in all directions, creating a ``pile-up'' in the middle.

Since OTVET is exact in the optically thin limit (aside from small distortions for light-front shapes, which are rarely important), artifacts only appear when the density is non-zero. The bottom row in Figure \ref{fig:arts} shows ionized bubbles around two unequal sources. The bubble around the much weaker source is not spherical, because the weaker source fills the optically thin contribution to the Eddington tensor from the stronger source \citep{Gnedin2001}.

These artifacts motivate one to search for better closure relations. One possible approach is to extend the moment hierarchy by considering closures of higher order. For example, recently \citet{Palanque2025} considered P$n$ closures, in which the tensor at the order $n$ is identically set to zero. Neither the P1 closure nor the P2 closure allows the single-source $1/r^2$ solutions, and thus only closures with $n\gg1$ can be considered. \citet{Palanque2025} demonstrated that the P9 closure offers an accurate solution to the two-source test. Unfortunately, the P$n$ closure requires $n(n+1)$ independent spatial fields (i.e.\ 90 for P9), which is not practical for large simulations.

In this paper, we only extend the hierarchy by one order and consider closures for the second-moment equation - the approximations for the heat tensor $Q^{ijk}$.

\section{Stability Analysis for Closures at the Second Order}

It is not difficult to invent a new closure, but any closure relation must, at the very least, be physically stable - the numerical implementation of a physically stable system can also be numerically unstable, but that is an issue with the numerical scheme, not with the physical equations. Let's now consider a base solution to the moment equations $(E_0, M_0^i, P_0^{ij})$ and a linear perturbation on top of it,
\[
    E=E_0+E_1, ~ M^i = M_0^i+M_1^i, ~ P^{ij} = P_0^{ij}+P_1^{ij}.
\]
Since the moment equations are themselves linear, both the base solution and the perturbation satisfy the moment equations independently (the perturbation satisfies equations with no source terms). Let us also consider the optically thin case ($\kappa=0$, as any absorption, being the damping term, is likely to suppress any instability rather than amplify it). As is usually done in linear stability analysis, the linear perturbation is taken to be a plane wave, $E_1, ... \propto e^{-i\omega t + i \vec{k}\vec{x}}$. Since $E_0, ...$ also depend on $\vec{x}$, we assume that $kL \gg 1$, where $L$ is the scale over which the base solution changes. The moment equations for the perturbations then reduce to
\begin{subequations}%
\begin{align}
    \omega E_1 & = c k^i M_1^i, \\
    \omega M_1^i & = c k^j P_1^{ij}, \\
    \omega P_1^{ij} & = c k^k Q_1^{ijk}.
\end{align}
\label{eq:master}
\end{subequations}
These are our ``master'' equations for the stability analysis presented below. The three equations can also be folded into a single characteristic equation by convolving all external indices with $k^i$,
\begin{equation}
    \omega^3 E_1 = c^3 k^i k^j k^k Q_1^{ijk}.
    \label{eq:c}
\end{equation}

\subsection{OTVET-like Closures at the Second Order}

One can easily generalize the OTVET closure to the second order,
\begin{equation}
    Q^{ijk} = E q^{ijk}
    \label{eq:oh2}
\end{equation}
with
\[
    q^{ijk} = \frac{1}{4\pi c E_{\rm OT}} \int d^3 x^\prime \frac{\dot{E}(\vec{x}^\prime)}{(\vec{x}-\vec{x}^\prime)^2} u^i u^j u^k
\]
and $E_{\rm OT}$ given by Equation (\ref{eq:eot}). Since $q^{ijk}$ is computed externally to the moment equations, it is not perturbed in our linear stability analysis, so $Q_1^{ijk} = E_1 q^{ijk}$ and the characteristic equation (\ref{eq:c}) becomes
\[
    \omega^3 = c^3 k^i k^j k^k q^{ijk}.
\]
Irrespective of the value for $q^{ijk}$, this equation has 1 real and 2 complex roots with non-zero imaginary parts, which makes the second-order OTVET-like closure (\ref{eq:oh2}) - somewhat unexpectedly - absolutely unstable.

On a side note, the original, 1st order OTVET closure is absolutely stable:
\[
    \omega^2 E_1 = E_1 c^2 k^i k^j h^{ij},
\]
and since $h^{ij}$ is positive definite, both roots of this equation are always real. Moreover, \emph{any} higher order OTVET-like closure is also unstable. For any $n>2$,  equation
\[
    \omega^n = c^n k^i k^j ... T^{ij...}
\]
has $n$ roots uniformly distributed over a unit circle in the complex plane, so at least 1 root has a negative imaginary part.

However, at the second order, there are more than 1 way to introduce an optically thin closure. One can notice that the tensor $Q^{ijk}$ is symmetric over all its indices, and a convolution over any pair of indices returns the first moment $M^i$. Hence, one can introduce a new tensor $W^{ijk}$ as
\begin{equation}
    Q^{ijk} = \frac{1}{5}\left(M^i\delta^{jk}+M^j\delta^{ik}+M^k\delta^{ij}\right) + W^{ijk}.
    \label{eq:ow2}
\end{equation}
Tensor $W^{ijk}$ is an analog of the Weyl tensor in General Relativity - it is a part of $Q^{ijk}$, which is not reducible to lower rank moments. Just like in GR, it cannot be omitted - the closure with $W^{ijk}$ set to zero does not allow the $1/r^2$ point source solution.

The optically-thin Weyl-like closure is exact for the point-source solution,
\[
    W^{ijk} = E w^{ijk},
\]
with
\begin{align}
    w^{ijk} = \frac{1}{4\pi c E_{\rm OT}} & \int d^3 x^\prime \frac{\dot{E}(\vec{x}^\prime)}{(\vec{x}-\vec{x}^\prime)^2} \left[u^i u^j u^k\right. \nonumber \\ 
    & \left.-\frac{1}{5}\left(u^i\delta^{jk}+u^j\delta^{ik}+u^k\delta^{ij}\right)\right]. \nonumber
\end{align}
The characteristic equation for the closure (\ref{eq:ow2}) is
\[
    \omega^3 - 3 \omega c^2 k^2 = c^3 k^i k^j k^j w^{ijk}.
\]
This equation has real roots if and only if $|k^i k^j k^j w^{ijk}|/k^3 < 2/(5\sqrt{5})$. Since $w^{ijk}$ can vary between -1 and 1 depending on where the sources are in the computational domain, this stability condition will not hold in general - i.e., this scheme is also unstable.

Finally, one can introduce another flavor of second-order OTVET: 
\[
    Q^{ijk} = \frac{M^iM^jM^k}{M^2} + W_2^{ijk}.
\]
However, in the optically thin limit, $W_2^{ijk}$ is identically zero, so this form is a special case of a local closure considered below.

\subsection{Restricted Local Closures at the Second Order}

A generalization of a higher-order M1-like scheme is a closure that depends only on the local values of the first two moments (i.e.\ on $E$ and $M^i$, but not on $P^{ij}$) - we will call such a local closure ``restricted''. With such restriction, the most general form for $Q^{ijk}$ is still Equation (\ref{eq:m2}) with an arbitrary $A(f)$. The second-order Maximum Entropy closure is not restricted, however - the heat tensor also depends on $P^{ij}$ via equations (\ref{eq:me2}). For such non-restricted closures, there is no finite general form for the heat tensor $Q^{ijk}$, and hence they are not studied here.

The characteristic equation for linear perturbations, a quintic polynomial equation for $x=\omega/(ck)$, is presented in Appendix \ref{app:ce2}. That equation is too cumbersome to be easily analyzable, so here we only consider a special case of $\alpha=0$ ($\alpha$ is the cosine of the angle between $\vec{k}$ and $\vec{M}_0$). Equation(\ref{eq:capp}) then reduces to
\begin{align}
x\left[ 25 x^4\right. & \left.+ 5\bigl(4(A-1)+A'f\bigr)\,x^2\right. \nonumber \\
& \left.+ 3(A-1)\bigl((A-1)+A'f\bigr) \right]=0. \nonumber
\end{align}
After factoring out the trivial root $x=0$, it becomes a quadratic equation for $y=5x^2$,
\[
y^2
+ \bigl(4(A-1)+A'f\bigr)\,y
+ 3(A-1)\bigl((A-1)+A'f\bigr)
=0,
\]
with roots
\[
y_1 = 3(1-A)
\qquad
y_2 = 1 - A - f A'.
\]
For the closure relation to be stable, both $y$ should be non-negative. The first condition is assured if $A\leq 1$. The second is
\begin{equation}
    A + f A' = (Af)' \leq 1
    \label{eq:cond}
\end{equation}
for all values of $f$.

The additional condition that the closure relation allows the single source $1/r^2$ solution requires $A(1)=1$. If we integrate the second condition from some value of $f$ to 1, we find:
\[
    \int_f^1 (Af)' df = 1 - Af \leq \int_f^1 1 df = 1-f,
\]
or
\[
    A \geq 1.
\]
Thus, we conclude that \emph{all local second-order closures that do not include $P^{ij}$ are unstable} except the special case $A(f)\equiv 1$ (which we consider below). 

\begin{figure}[t]
\centering
\includegraphics[width=\columnwidth]{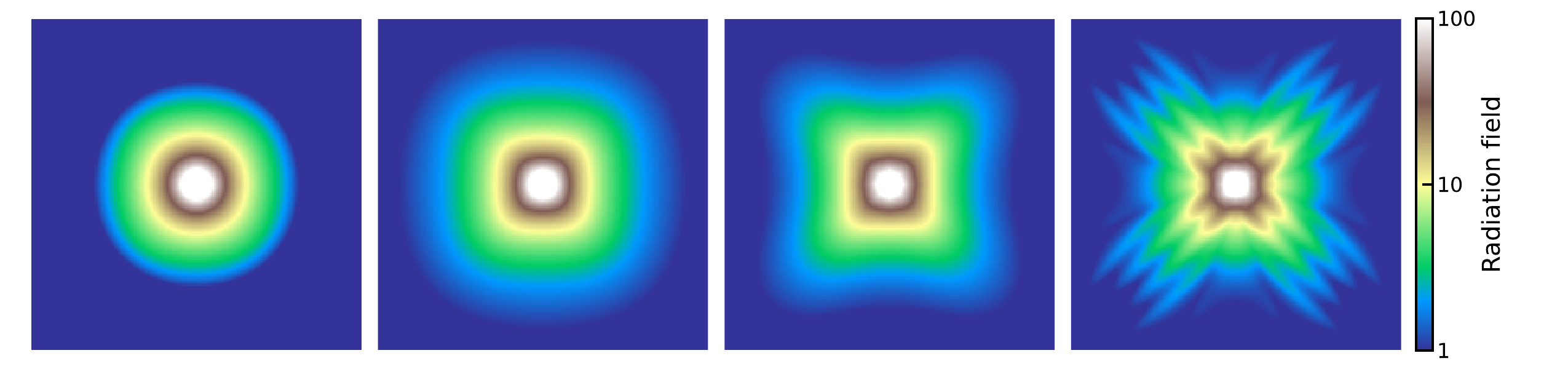}\newline%
\includegraphics[width=\columnwidth]{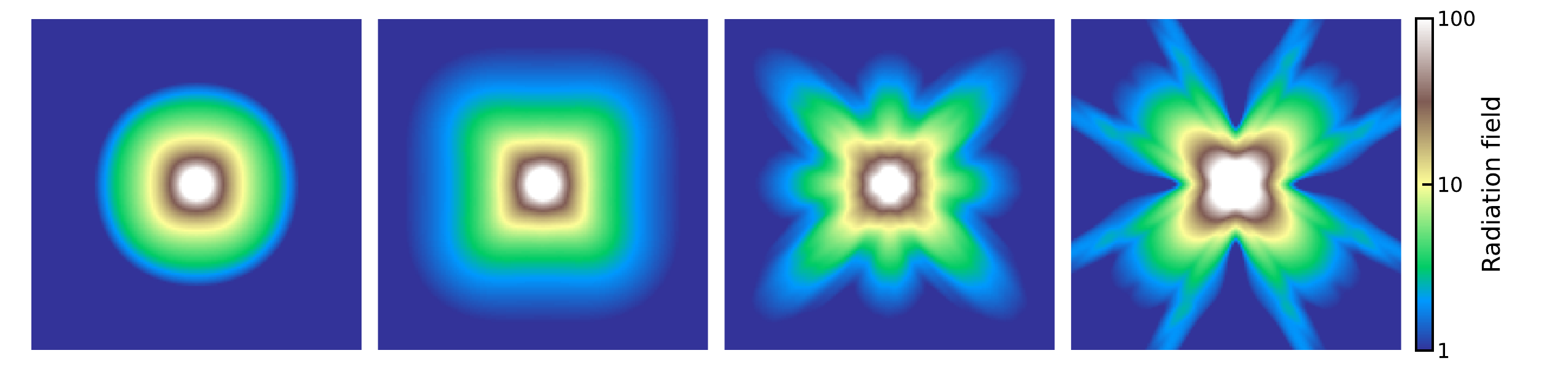}%
\caption{Radiation field (in arbitrary units) from a single isotropic source in the empty space at four snapshots. The top row shows the second-order closure with $A(f)=f$ and the bottom row is for $A(f)=f^3$. The instability develops faster in the bottom row than in the top row, as expected.}
\label{fig:fn}
\end{figure}

As an illustration, we implemented a numerical solver with $A(f)=f^n$ closures. Figure \ref{fig:fn} shows the development of the instability for a test with a single source in the empty space. It is important to note that the instability does not imply that, say, the radiation energy density becomes infinite. The moment equations ensure photon energy density and flux conservation, so even with the unstable closure, the total photon energy density is conserved, and the spherically averaged solution around a point source is still $L/(4\pi c r^2)$. However, the angular distribution of the radiation becomes non-spherical with time due to instability. For $A(f)=f^n$, the value of $(Af)'$ at $f=1$ is $n+1$, hence one expects the instability to grow faster in the $n=3$ case than in the $n=1$ case. 

\vspace{1cm}
\subsection{Special Case $A(f)\equiv 1$}

In this special case the characteristic equation (\ref{eq:capp}) reduces to
\[
    x(x^2-\alpha^2)^2 = 0,
\]
with 5 real roots. Hence, this closure is linearly stable in the limit of large $k$. A full solution for linear perturbations for the case of the $1/r^2$ point source solution is presented in Appendix \ref{app:a=1}.

\begin{figure}[t]
\centering
\includegraphics[width=\columnwidth]{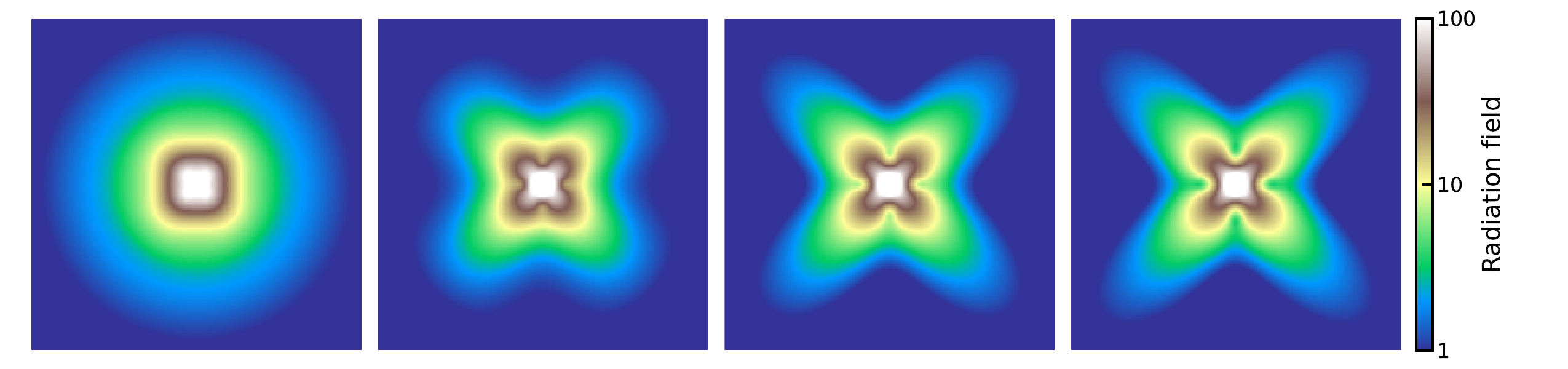}\newline%
\includegraphics[width=\columnwidth]{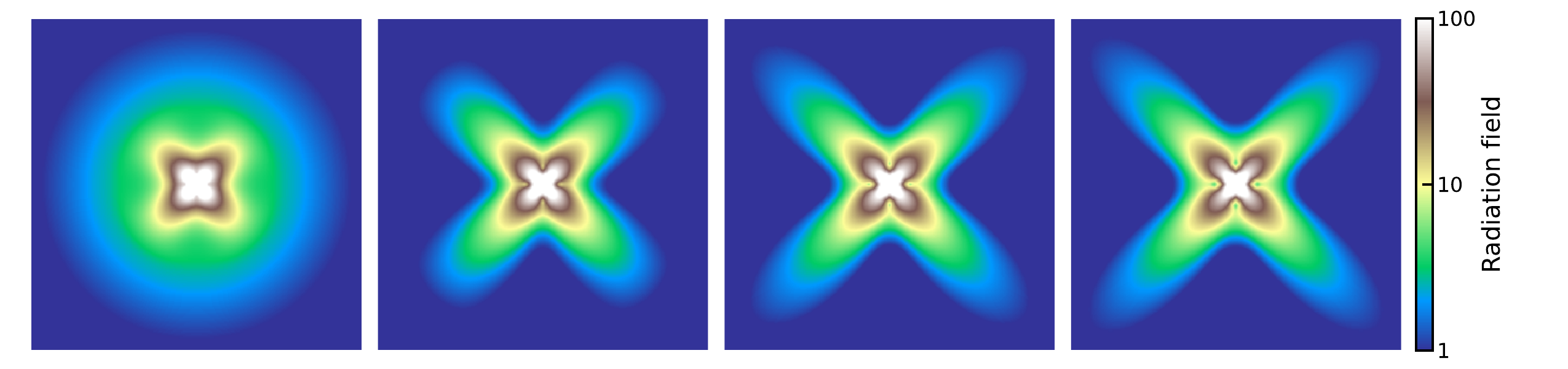}\newline%
\includegraphics[width=\columnwidth]{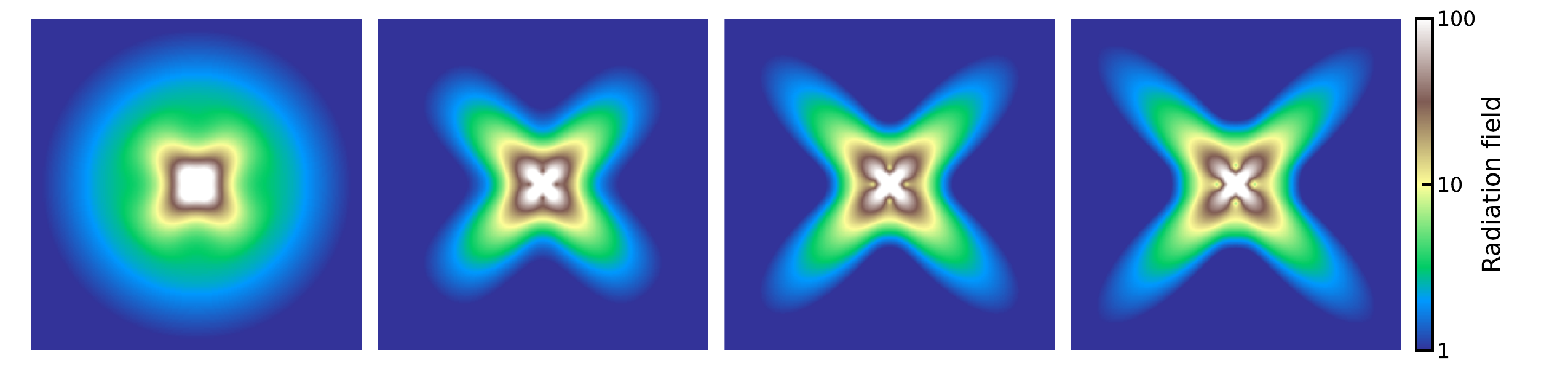}%
\caption{Radiation field (in arbitrary units) from a single isotropic source in the empty space at four snapshots linearly spaced in time for the special case $A(f)\equiv 1$ closure. Simulations start with the exact solution $E=L/(4\pi c r^2)$ to avoid any dynamical effect of a light front propagating from an instantly switched-on source. The top and middle panels show tests with point sources and effective resolution of $512^3$ and $1024^3$, respectively (only 1/6 of the simulation volume is shown in the image). The bottom row shows the same test as in the middle row, with the source being spread in a Gaussian cloud with $\sigma$ equal to 3 cell sizes. Despite being linearly stable, the closure does develop non-spherical structures in the solution.}
\label{fig:a1}
\end{figure}

Figure \ref{fig:a1} shows the single-source test for this special case. Despite being formally linearly stable, the solution robustly develops the non-spherical features. These features are \emph{not} numerical artifacts: they persist at different numerical resolutions, time-steps, and box sizes (i.e.\ they are not artifacts of boundary conditions). They also develop at the same rate if, instead of switching a source at some initial time, we start with the established $1/r^2$ distribution already. 

Finally, a point source deposited on a regular grid may imprint the grid in the solution. To check for such a possibility, we also replaced the point source with a Gaussian-shaped spherical source with a width of 3 cells. The same asymmetries appear in this case as well, and at the same rate. In addition, we also tried different numerical schemes to make sure the developing asymmetry is not a numerical artifact of our specific implementation. We tried both, a higher, fourth-order finite difference scheme and a second-order scheme with a different stencil for finite differencing. The default scheme used here is the one written out explicitly in \citet{Gnedin2025}, and uses the most common of second-order finite differencing ($df/dx|_i \approx (f_{i+1}-f_{i-1})/(2\Delta x)$). We also tried a different finite difference representation that used the full 26-neighbor stencil. The numerical solution remained the same in both cases.

\begin{figure}[t]
\centering
\includegraphics[width=\columnwidth]{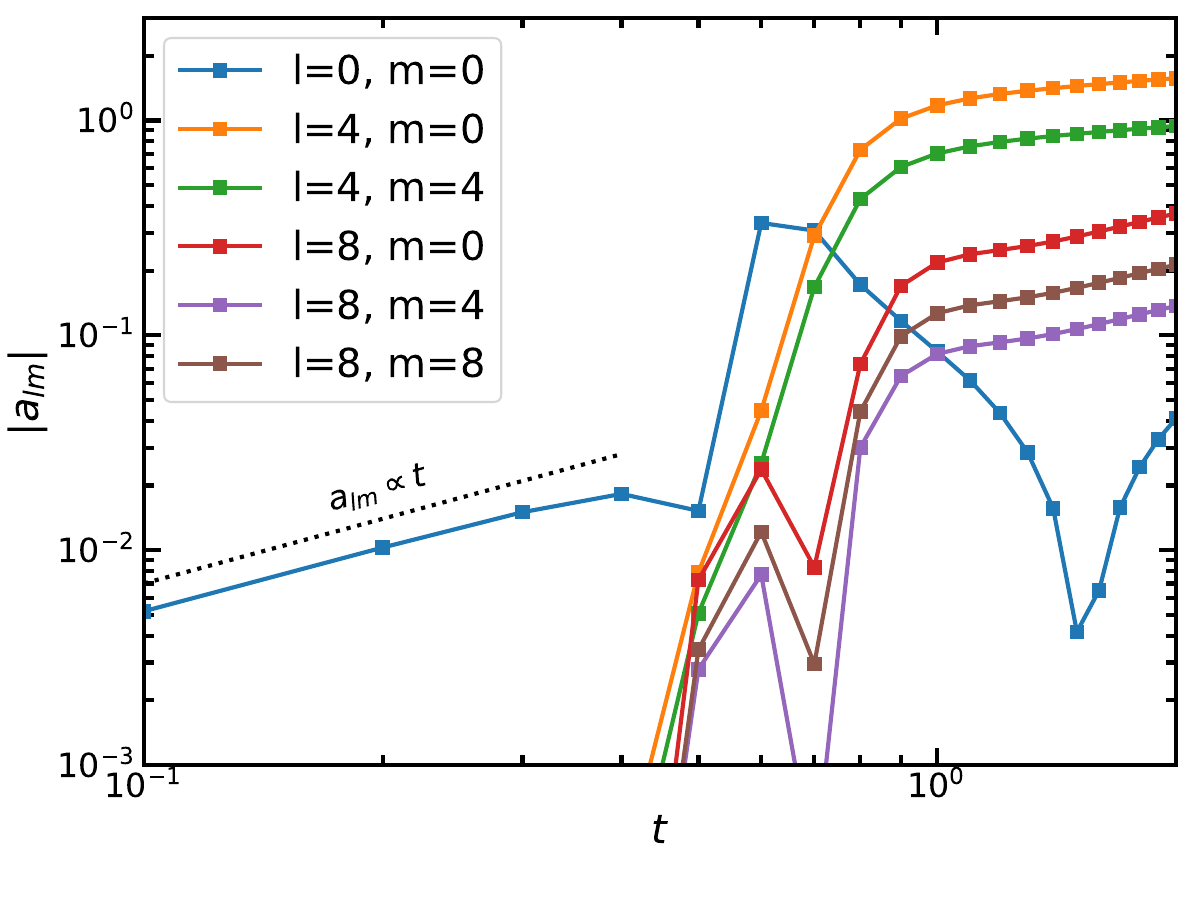}%
\caption{Evolution of non-trivial multipoles for the numerical solution of the $A(f)\equiv 1$ test at a distance of 0.1 of the box size from the source. Initially, the spherical symmetry is preserved, but the deviation from the exact solution increases linearly with time until it enters the non-linear regime, at which point the hexadecapole and later higher overtones appear.}
\label{fig:alm}
\end{figure}

In short, we have checked all known sources of numerical artifacts and found the developed asymmetries to be fully robust against them. We, therefore, believe that they are indeed the correct solution of the moment hierarchy with the $A(f)\equiv 1$ closure. To explore this unexpected behavior further, we show in Figure \ref{fig:alm} the evolution of all nontrivial multipoles of the numerical solution on a sphere at radius 0.1 of the box size from the source. The deviations from the exact solution start as a monopole, and increase linearly with time (i.e., logarithmically slower compared to the linearly unstable behavior of $e^{{\rm Im}\omega ct}$). Eventually, the deviations become nonlinear and generate $l=4$ perturbations and their overtones. 

This behavior can be plausibly understood from the full linear perturbation analysis presented in Appendix \ref{app:a=1}. For $l=0$, only the scalar harmonic is present, and the static mode $\omega=0$ is constant. Normally, this would indicate that the solution is stable, since the amplitude of the perturbation should be sufficiently small. However, this is only true for the perturbation to $M^i$. Since the photon energy density $E$  satisfies Equation (\ref{eq:moms}a), the perturbation to $E$ grows linearly with time. This growth is not numerical (in the sense that it is encoded in the equations themselves, not in their finite difference form), and hence the numerical solution is stably recovering the physical behavior dictated by the equations.

\section{Conclusions}

A general analysis of the stability of second-order local closure schemes is beyond the scope of this work, as there are infinitely many possible forms for the tensor $Q^{ijk}$ that include $P^{ij}$. Hence, we only considered the ``restricted" local closure, in which $Q^{ijk}$ depends only on $E$ and $M^i$. Such closures are direct generalizations of the commonly used M1 closure. We have shown that all such closures are physically unstable. In addition, two possible second-order generalizations of the alternative first-order closure, OTVET, are also unstable.

Hence, the task of developing second-order closure relations is significantly more challenging than it may initially appear. All possible stable local closures (if they exist at all) must depend on the radiation pressure tensor $P^{ij}$ in a non-trivial way. Namely, since the instability for the restricted closures appears even for small departures of $f\equiv M/E$ from unity, all stable closures must explicitly depend on $P^{ij}$ even when radiation is freely streaming in a single direction, $M \approx E$.

\section*{Acknowledgments}

The first draft of the introduction and most of the analytical calculations in this paper were performed by ChatGPT. This work was supported by Fermi Forward Discovery Group, LLC under Contract No.\ 9243024CSC000002 with the U.S. Department of Energy, Office of Science, Office of High Energy Physics. We also acknowledge the support from the University of Chicago’s Research Computing Center, where the largest simulations used in this work were completed. The material in this manuscript is based upon work supported by NASA under award No. 80NSSC25K7009.

\appendix

\section{A. Characteristic Equation for the Restricted Local Second-Order Closure}
\label{app:ce2}

The full derivation of this equation is documented in the ChatGPT conversation at this url: \url{https://drive.google.com/open?id=1m0iM5tuQjFLIs1aQBXFVqIL-qXn_Iyg-&usp=drive_fs}.

If $x=\omega/(ck)$, then the characteristic equation is a quintic polynomial equation for $x$, 
\begin{equation}   
25 x^5
-5\Big(10A\,\alpha^2-4A+3A'f\,\alpha^2-A'f+4\Big)x^3
+5A' f^2\,\alpha(5\alpha^2-3)x^2
+\mathcal{C}_1\,x
-\;A'f^2\,\alpha\Big(5A\alpha^4+3A+5\alpha^2-3\Big)
=0,
\label{eq:capp}
\end{equation}
where $\alpha$ is the cosine of the angle between vectors $\vec{k}$ and $\vec{M}_0$ ($\vec{k}\vec{M}_0 = kM_0\alpha$), $A^\prime = dA/df$, and 
\[
\mathcal{C}_1=
5A^2\alpha^4 + 3A^2
+5AA'f\,\alpha^4 + 3AA'f
+20A\alpha^4 - 6A
-10A'f\,\alpha^4 + 15A'f\,\alpha^2 - 3A'f
+3.
\]

\section{B. Full Linear Stability Analysis for the Special Case $A\equiv 1$.}
\label{app:a=1}

In a general case where one cannot assume that the linear perturbation depends on $\vec{x}$ as $e^{i\vec{k}\vec{x}}$, one can still perform the full linear analysis of the special case $A\equiv 1$ for the case of a single point source. For this closure, one can obtain a single equation for $\vec{M}$ by combining 1st and second moment equations:
\[
    \frac{1}{c^2}\frac{\partial^2 M^i}{\partial t^2} = \frac{\partial^2}{\partial x^j x^k} \frac{M^iM^jM^k}{M^2}.
\]

If $M_0^i \propto r^i/r^3$, where $r$ is the spherical radius, then the perturbation $M_1^i$ can be written as 
\[
    M_1^i = \frac{e^{-i c\omega t}}{r^2}\left(n^i y(r,\Omega) + t^i(r,\Omega)\right),
\]
where $n^i\equiv r^i/r$ and $t^i$ is strictly transverse, $n^i t^i=0$. 

Perturbations $y(r,\Omega)$ and $t^i(r,\Omega)$ can further be expanded in scalar and vector spherical harmonics respectively, 
\[
y(r,\Omega)
=
\sum_{l=0}^{\infty}
\sum_{m=-l}^{l}
y_{l m}(r)\,Y_{l m}(\Omega),
\]
\[
t^i(r,\Omega)
=
\sum_{l=1}^{\infty}
\sum_{m=-l}^{l}
\Big[
a_{l m}(r)\,\Psi_{l m}^i(\Omega)
+
b_{l m}(r)\,\Phi_{l m}^i(\Omega)
\Big],
\]
where
\[
\Psi_{l m}^i(\Omega)
:=
\nabla_S^i Y_{l m}(\Omega),
\]
and 
\[
\Phi_{l m}^i(\Omega)
:=
\left(n \times \nabla_S Y_{l m}\right)^i,
\]
and 
\[
\nabla_S^i
:=
\left(\delta^{ij}-n^i n^j\right)\partial_j
\]
is a surface gradient on a unit sphere.

The full derivation of equations for linear perturbations is documented in the ChatGPT conversation at this url: \url{https://drive.google.com/open?id=113KeMmG3_pBlf4XydqbWPzd2O-Rh-zSy&usp=drive_fs}. The toroidal component $b$ decouples,
\[
b'' + \frac{2}{r} b'
+ \left(\omega^2 - \frac{l(l+1)-1}{r^2}\right) b
= 0.
\]
The radial and poloidal components form a coupled system,
\[
y'' + \frac{2}{r} y'
+ \left(\omega^2 - \frac{6}{r^2}\right) y
- \frac{2l(l+1)}{r}\left(a' + \frac{a}{r}\right)
= 0 ,
\]
\[
a'' + \frac{2}{r} a'
+ \left(\omega^2 - \frac{l(l+1)+1}{r^2}\right) a
+ \frac{2}{r}\left(y' - \frac{2y}{r}\right)
= 0 .
\]
The solution of this coupled system of equations is
\[
\frac{d}{dr}\!\left(\frac{y(r)}{r^{2}}\right)
=\frac{C_+\,j_{l+1}(\omega r)+C_-\,j_{l-1}(\omega r)}{2r^{4}},
\]
\[
a(r)=
=\frac{C_-\,j_{l-1}(\omega r)-C_+\,j_{l+1}(\omega r)}{2\sqrt{l(l+1)}},
\]
and
\[
b(r)=B\,j_{\lambda}(\omega r),
\]
where $j_l(x)$ is a spherical Bessel function, $C_\pm$ and $B$ are integration constants, and $\lambda(\lambda+1)=l(l+1)-1$.

\bibliographystyle{mnras}
\bibliography{main}

\end{document}